\documentclass[referee]{raa}            % referee version: for submission

\usepackage{graphicx,times}             %for PS/EPS graphics inclusion, new
\usepackage{natbib}
\usepackage{amssymb,amsmath}
\bibpunct{(}{)}{;}{a}{}{,}
\usepackage{float}

\usepackage[pagebackref=true]{hyperref}

\begin{document}

  \title{ LEO Satellite Track Correction for CSST Multi-Band Imaging Data 
  }
%   \subtitle{I. Place Your Subtitle Here}

   \volnopage{Vol.0 (20xx) No.0, 000--000}      %%preserved for Editor. DOn't remove!
   \setcounter{page}{1}          %%starting page, preserved for Editor. DOn't remove!

   \author{Huai-Jin Tang %(唐怀金) %% Put your Chinese name in "( )" if you like. Note to open line 11 "\usepackage[UTF8]{ctex}"
      \inst{1,2}
   \and Xiao-Lei Meng
      \inst{1}
   \and Hu Zhan
      \inst{1,3}
    \and Guo-Liang Li
      \inst{4}  
    \and Cheng-Liang Wei
      \inst{4}  
   \and Xian-Min Meng
      \inst{1}
    \and Xi-Yang Fu 
      \inst{1}
   \and You-Hua Xu
      \inst{1}      
    % 孟宪民，伏西洋，许优华,李国亮，csst其他工作人员
      }
%% Here is an example of three authors come from different institutes.
%% For single author or all the authors from an institute, use "\inst{}" only

   \institute{Key Laboratory of Space Astronomy and Technology, National Astronomical Observatories, Chinese Academy of Sciences, Beijing, 100101, China; {\it tanghj@nao.cas.cn}\\
%% Please give the E-mail address of the author, to whom future correspondence and
%% offprint requests will be sent.
        \and
             University of Chinese Academy of Sciences (UCAS), Beijing 100049, China\\
        \and
             Kavli Institute for Astronomy and Astrophysics, Peking University, Beijing, 100871, China\\
                     \and
             Purple Mountain Observatory, Chinese Academy of Sciences, Nanjing 210023, China\\
\vs\no
   {\small Received 20xx month day; accepted 20xx month day}}

\abstract{
Low Earth Orbit satellite (LEOsat) mega-constellations are considered to be an unavoidable source of contamination for survey observations to be carried out by the China Space Station Telescope (CSST) over the next decade. This study reconstructs satellite trail profiles based on simulated parameters, including brightness levels and orbital altitudes, in combination with multi-band simulated images. Compared to our previous work, the simulated images in this study more accurately replicate the realistic observational conditions of CSST and extend beyond single-band analysis. Variations in LEOsat trail brightness, source brightness, background noise, and source density across different bands result in differing levels of accuracy in trail reconstruction and subsequently affect the reliability of photometric measurements. The reconstructed trail profiles are subsequently applied to correct the contaminated regions. Simulation results reveal varying levels of contamination effects across different bands following LEOsat trail correction, including both reconstruction and subtraction. To evaluate the effectiveness of the correction, we quantified the fraction of affected sources using two metrics: (1) magnitude errors greater than 0.01 mag attributable to LEOsats, and (2) LEOsat-induced noise exceeding 10\% of other noise contributions. Following trail repair, the analysis reveals a reduction of over 50\% in the fraction of affected sources in the NUV band for both 550 km and 1200 km altitudes, assuming a maximum brightness of 7 in the V band. In the $i$ band, the reduction exceeds 30\%. The degree of improvement varies across spectral bands, and depends on both satellite altitude and the adopted brightness model.
}
   \authorrunning{H.-J. Tang, X.-L. Meng, H. Zhan }            %author_head in even pages
   \titlerunning{LEO Satellite Track Correction for CSST Multi-Band Imaging Data  }  % title_head in odd pages
   \maketitle
%% The author head (on even pages) and the title head (on odd pages) will be
%% automatically extracted from \author{} and \title{}. Whenever the title is too long,
%% you will be asked to supply a shorter one by inserting either \authorrunning{} or
%% \titlerunning{} before \maketitle. Anyway, you can specify your own heads.
%%
%%
%% Note: In the following text body of your manuscript, please note several differences from
%%       other major journals:
%% (1) \subsection{Please Capitalize the First Letter of Each Notional Word in Subsection Title}
%% (2) Please Capitalize the First Letter of Each Notional Word in all tables' captions
%
%________________________________________________ sections below
%

\section{Introduction}
With the rapid deployment of low Earth orbit satellite (LEOsat) mega-constellations such as Starlink, OneWeb, and others, the number of operational satellites in near-Earth space is projected to rise to $10^5$ over the next decade \citep{Venkatesan2020, Walker2020, Rawls2021, Bassa2022}. These satellites typically operate at altitudes ranging from 300 km to 1200 km and adopt large orbital inclinations to achieve near-global coverage \citep{Walker1984, Walker2020, Bassa2022}. The increasing density and wide spatial distribution of these constellations have introduced significant challenges for wide-field astronomical surveys. In particular, LEOsat trails can severely affect transient event monitoring, crosstalk phenomena, and multi-wavelength coordinated observations, degrading source detection completeness and compromising the accuracy of photometric and morphological measurements \citep{ Laher2014,Tyson2020, Walker2020, Mroz2022}. Mitigating the impact of LEOsats on astronomical observations has therefore become an imperative task of the astronomical community. Current mitigation strategies include encouraging satellite operators to modify surface reflectivity properties, implement optimized attitude control schemes, improve the precision of satellite ephemerides, classify satellite types, and refine satellite brightness models \citep{McDowell2020,  Krantz2022,Fankhauser2023, Kruk2023, Lu2024}. Accordingly, systematic and quantitative analyses of LEOsat trail characteristics, along with the development of effective removal and correction techniques, are essential to preserve the scientific integrity of wide-field astronomical surveys.

The impact of LEOsat interference on multi-band astronomical observations is primarily governed by the satellites’ apparent brightness as observed by the telescope. At present, the brightness of LEOsats remains excessively high for ground-based observers. Observational studies indicate that Starlink V1 satellites, including the Darksat and VisorSat variants, typically exhibit apparent magnitudes between 5.5 and 6 AB at an operational altitude of approximately 550 km \citep{Walker2020, Tregloan2020, Mallama2021, Tregloan2021, Halferty2022}, substantially exceeding the background sky brightness in ground-based astronomical images \citep{Walker2020}. The Satellite Constellations 1 (SATCON1) report recommends that satellite brightness be reduced to no brighter than 7 AB magnitudes to minimize their impact on astronomical observations \citep{Walker2020}. In response to these concerns, the latest generation of Starlink V2 satellites incorporates design modifications specifically intended to reduce their apparent brightness. Simulation results have demonstrated that these measures are effective in mitigating contamination from satellite trails in astronomical images \citep{Kandula2025}. Furthermore, extensive multi-band observations of LEOsats have yielded substantial datasets and analysis  results, contributing valuable insights into their photometric properties and temporal behaviors \citep{Campbell2019, Hossein2023, Zhi2024}.

Effectively mitigating the impact of LEOsats on multi-band space-based astronomical surveys is essential for preserving data quality. The number of satellites entering a telescope’s field of view (FoV) depends on several factors, including the FoV area, exposure time, the spatial density and velocity of satellites, and the observing geometry \citep{Tyson2020, McDowell2020, Hainaut2020, Kruk2023}. Differences in orbital altitude further influence the frequency of satellite crossings, with higher-altitude satellites typically traversing the FoV more often than those at lower altitudes \citep{Walker2020, Bassa2022, Kruk2023}. Although space-based telescopes generally have smaller apertures, they benefit from a stable orbital environment, a consistent point spread function (PSF) free from atmospheric turbulence (\citealt{Walker2020, Hasan2022}), and relatively moderate trail brightness, offering favorable conditions for identifying and correcting satellite contamination.

As a representative of the next generation of advanced space-based observatories, the China Space Station Telescope (CSST) is designed to perform multi-band imaging across seven photometric filters (NUV, $u$, $g$, $r$, $i$, $z$, $y$) and three slitless spectroscopic filters (GU, GI, GV) \citep{Zhan2006, Zhan2011, Cao2018, Gong2019, Zhan2021, Cao2022, Li2023, Miao2023}. Operating in a LEO at an altitude of approximately 400 km, the CSST features a 2-meter primary mirror and a wide FoV of 1.1 square degrees. The focal plane is equipped with an array of thirty CCD detectors. The mission of CSST main survey includes a wide-area imaging survey covering 17,500 square degrees, with each field observed with a total exposure time of 2 $\times$ 150 seconds. Our work primarily focuses on investigating the impact of LEOsats on CSST observations.

In our previous studies, the COSMOS data (Cosmic Evolution Survey) from the band of HST F814W were used to simulate the single-band performance of CSST in the $i$-band \citep{Tang2025}. Based on the mock data, we investigate the impact of LEOsat orbital altitude and brightness on CSST’s seven-band imaging performance in this work. We evaluate their effects on data quality and assess the effectiveness of trail removal techniques. Moreover, we utilize simulated imaging datasets developed by the CSST collaboration \citep{Wei2025} to explore a trail reconstruction-based approach, wherein the morphological profiles and brightness distributions of satellite trails are modeled and subtracted from contaminated images. This methodology provides a foundation for developing optimized interference mitigation strategies for multi-band CSST data processing. Compared to commonly used trail mitigation techniques—such as masking, inpainting, and PCA-based methods—our reconstruction–subtraction approach offers notable advantages, particularly for space-based telescope imaging and the CSST simulation framework. Masking typically targets only the central high-signal regions of trails, often leaving edge areas insufficiently treated and prone to residual contamination. Inpainting methods can fill missing pixels but generally fall short of the precision required for accurate photometric and morphological measurements. PCA-based methods depend on multi-epoch statistical information, limiting their use in single-exposure images. In contrast, our approach provides greater flexibility and accuracy, enabling more effective recovery of astronomical information in trail-contaminated regions and enhancing both data utilization and measurement reliability.

This paper is organized as follows. Section 2 provides a detailed description of the simulation procedures, including the image simulation workflow, LEOsat characterization, and the computational methods. Section 3 presents the results of the multi-band simulations, demonstrating the effects of satellite trail reconstruction and removal, and evaluating the performance and potential of various trail correction methods. Finally, Section 4 summarizes the overall findings, offers conclusions, and discusses future directions for optimizing satellite trail mitigation techniques in the processing of CSST multi-band observational data.

\section{Data and Method}

\subsection{Image simulation}
The image simulations employed in this study are based on data products and simulation tools developed by the CSST scientific data processing and analysis system \citep{Wei2025}. During the generation of CSST multi-band simulated images, the simulation is based on observational parameters and the True Universe catalog. The stellar catalog is generated using the Galaxia tool based on the Besançon model of the Milky Way \citep{Sharma2011, Robin2003}. The galaxy catalog is constructed from the high-resolution JiuTian (JT1G) cosmological N-body simulation, combined with a semi-analytical galaxy formation model \citep{Luo2016}. Galaxy sizes are determined using the mass–size relation of SDSS galaxies \citep{Zhang&Yang2019}, with an additional redshift-dependent scaling factor to account for cosmic size evolution. The spectral energy distributions (SEDs) of galaxies are generated using the STARDUSTER deep learning model \citep{Qiu&Kang2022}. The quasar catalog is generated with SIMQSO, based on observed luminosity functions, and quasars are matched to active galactic nuclei (AGN) in the galaxy catalog \citep{McGreer2021}. In addition, weak gravitational lensing effects are modeled by constructing a past light cone from the JT1G simulation and performing full-sky ray-tracing, which are consistent with theoretical predictions \citep{Wei2018}. The simulated images incorporate various realistic instrumental and observational effects, including background noise, detector noise, hot and warm pixels, cosmic rays, and other noise sources \citep{Wei2025}. Some of these artifacts are flagged by the CSST data processing pipeline to assess their impact on single-exposure observations. However, some residual cosmic rays and small bright pixels remain unflagged. These residual features must be taken into account when identifying bright sources and reconstructing LEOsat trail profiles. In this simulation, a sky region of approximately 2 $\times$ 2 square degrees is selected.

\subsection{LEOsat parameters}

In studying the impact of LEOsat interference on multi-band astronomical observations, factors such as orbital altitude, apparent brightness, and spectral reflectance properties play a critical role. This study compares the effects of LEOsats under different orbital altitudes (550 km and 1200 km, 53 degrees orbital inclination) and brightness levels. The simulation adopts two representative orbital altitudes: 550 km, a typical Starlink orbit known to affect CSST observations, and 1200 km, which approaches the upper limit of LEO. Recent advancements in precise Bidirectional Reflectance Distribution Function (BRDF) models have greatly improved satellite brightness predictions, allowing for more accurate assessments of their impact on astronomical imaging \citep{Fankhauser2023, Lu2024}. In this work, a simplified apparent magnitude model is adopted as a first-order approximation. The V-band apparent magnitude of a LEOsat (V$_{\rm sat}$) is estimated following the approaches of \citet{Hainaut2020} and \citet{Walker2020}.

The maximum satellite brightness (V$_{\rm M}$) is set based on recommended thresholds, with a typical value of 7 magnitudes at an orbital altitude of 550 km \citep{Walker2020}. To investigate the contamination effects of satellites at different brightness levels, we compare cases with minimum apparent magnitudes (V-band V$_{\rm M}$ values of 5 and 7). The solar phase angle contribution is expressed as $v_{\alpha} = (1+\cos(\alpha))/2$ \citep{Hainaut2020}. Additionally, when converting magnitudes for space-based observations, a V-band atmospheric attenuation of 0.2 magnitudes is subtracted to account for the absence of atmospheric extinction \citep{Patat2011}. As an example, the satellite magnitude in the NUV band ($M_{\rm NUV, \rm sat}$) is calculated using Equation 1.

\begin{equation}
   M_{\rm NUV, \rm sat}=1.4+ {\rm V}_{\rm M} + 5 {\rm log_{10}}(\frac{D_{\rm ist}}{550\mbox{ } \rm km}) - 2.5 {\rm log_{10}}(v_{\alpha}) - 0.2
   \label{eq:quadratic}
\end{equation}

Here, $D_{\rm ist}$ is the distance of the satellite to the telescope. The difference between the minimum apparent magnitude in the V-band (V$_{\rm M}$) and those in the CSST photometric bands is given by NUV – V$_{\rm M}$ = 1.4 mag, and ($u$, $g$, $r$, $i$, $z$, $y$) – V$_{\rm M}$= (0.56, –0.40, –0.32, –0.064, –0.10, –0.10) mag, respectively. The NUV band is a representative and important observational band for characterizing satellite trails. Considering the photon counts, LEOsat trails in the NUV band are significantly fainter compared to those in other bands. A comparison with the relatively brighter $i$-band will be presented later.

The next step involves converting the simulated apparent V-band magnitudes of satellites to the corresponding brightness values in each observational band of CSST. Following the approach of previous studies \citep{Tyson2020}, a solar-like spectral energy distribution (SED) is adopted to represent the SED of LEOsats. To compute the brightness of satellite trails in each band, the system throughput for CSST’s photometric filters is applied \citep{Cao2018, Zhan2021, Cao2022}. Taking the system efficiency into account, the conversion factors from V-band photon counts to electron counts in the seven photometric imaging bands (NUV, $u$, $g$, $r$, $i$, $z$, $y$) are determined to be 0.0349, 0.148, 0.859, 1.08, 1.06, 0.639, and 0.184, respectively.

Then, the simulation involves constructing the morphological profiles of multi-band satellite trails. To achieve this, the CSST optical system is simulated to obtain the off-focus PSF for each photometric band. The off-focus PSFs are subsequently convolved with modeled satellite images, assuming a physical size of 2 meters \citep{Tyson2020}, to generate one-dimensional cross-sectional profiles of the LEOsat trails. Due to the out-of-focus imaging design of the system, the apparent sizes of satellite trails remain nearly uniform across different bands. Consequently, the full width at half maximum (FWHM) of the trails is measured to be highly consistent among all bands, with the specific values summarized in Table 1.

\begin{table}[htbp] 
\center
\caption{The FWHM (pixel) of the off-focus LEOsat trails at various distances in different bands  }
\begin{tabular}{lllllllll}
\hline
&Distance  &NUV &$u$   &$g$  &$r$  &$i$  &$z$  &$y$   \\
\hline
&100 km &56.1  &56.3  &56.4  &56.5  &56.6   &56.7  &56.8 \\
&200 km &28.2 &28.3  &28.4 &28.5 &28.5  &28.6 &28.6  \\
&400 km &14.2 &14.2  &14.3 &14.4 &14.5  &14.5 &14.6 \\
&600 km &9.2  &9.2  &9.3  &9.4 &9.5 &9.5 &9.6  \\
&800 km &7.1 &7.2   &7.2 &7.3  &7.3   &7.4  &7.4   \\
&1000 km &5.6 &5.6  &5.7 &5.7 &5.8  &5.8   &5.8 \\
&2000 km &2.6  &2.7  &2.7 &2.8 &2.8  &2.8   &2.8 \\
\hline
\end{tabular}
%\label{Table:NBstart}
\end{table}

\subsection{LEOsat simulation}

In the LEOsat orbital simulation, we adopt the parameter settings described above and distribute a population of $10^4$ LEOsats uniformly across 100 orbital planes at a given altitude. This setup is combined with the simulated orbit and observational schedule of the CSST. The trail brightness is determined by the apparent magnitude of the satellites and their angular velocity as they transit the CSST FoV, while the trail profile is influenced by the satellite’s distance. The detailed simulation procedures and results—including the number of satellites entering the FoV, their distances, apparent magnitudes, trail brightness, and corresponding trail lengths on the detector—are provided in Tang et al. (2025) \citep{Tang2025}. Assuming a population of $10^5$ LEOsats distributed across various altitudes, and scaling from results obtained with $10^4$ simulated LEOsats, the expected length of bright trails across a single CCD is approximately 1800 pixels for satellites at an altitude of 550 km, and around 4865 pixels for those at 1200 km.

Image simulations are conducted using the LEOsat trail parameters obtained from orbital simulations. The mean angular speed of LEOsats relative to the CSST is approximately 1.26 degrees per second at an altitude of 550 km, and 0.35 degrees per second at 1200 km. The mean distance of LEOsats relative to the CSST is approximately 492.4 km at an orbital altitude of 550 km, and 1588.1 km at 1200 km \citep{Tang2025}. In most cases, a LEOsat traverses an entire CCD during its transit across the CSST FoV. When configuring the length and orientation of LEOsat trails, shorter trails produce lower cumulative signal-to-noise ratios (SNRs) but contaminate fewer pixels, while longer trails result in higher cumulative SNRs and affect a larger number of pixels. To replicate realistic observational conditions, satellite trails are randomly distributed across the detector, with their lengths, orientations, and entry and exit positions assigned along straight-line paths from one edge to another. This randomization ensures that the statistical distributions of trail lengths and orientations closely resemble those observed during actual satellite transits, providing a representative model of contamination patterns. In the simulated images, each LEOsat trail is added as the product of its flux and the trail’s cross-sectional profile, along with the corresponding photon noise. Despite subsequent trail removal procedures, imperfect reconstructions often leave behind residual artifacts. The combined effects of photon noise and reconstruction residuals are quantitatively assessed to evaluate their impact on measurement accuracy and to determine the overall effectiveness of the trail correction process. Figure 1 shows a bright LEOsat trail crossing a simulated $i$-band image.

\begin{figure}
\centering
\includegraphics[width=0.95\linewidth]{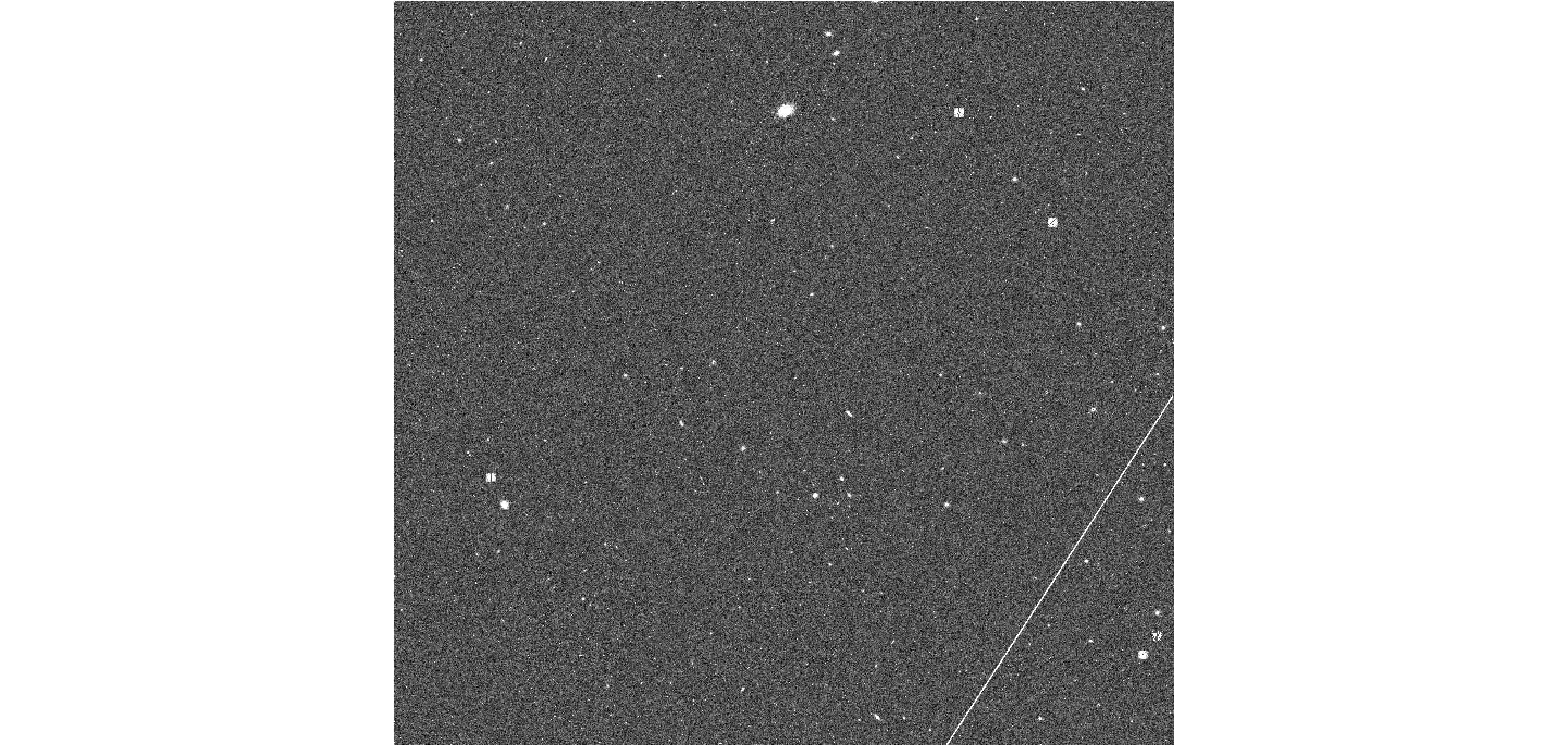}
\caption{The figure illustrates the trail of a LEOsat across a simulated $i$-band CCD (9216 $\times$ 9232 pixels), which covers an area of 129.4 square arcminutes.}
\end{figure}

\subsection{Method}

The trail profile used in our reconstruction method is obtained from simulated satellite trails and empirically fitted to reconstruct the profile. Specifically, we stack the simulated trails along their direction to enhance the SNR of the trail profile, which is then empirically fitted to generate a template for subtraction. Overlapping trails are extremely rare given the typical exposure times of CSST; thus, no special treatment has been implemented in the current version. The detailed computational procedure is described below.

In the LEOsat trail reconstruction and correction procedure, the initial step focuses on repairing pixels flagged by the data processing pipeline. This is achieved by applying a median filter to replace the affected pixel values based on the surrounding pixel information. When bright sources overlap with satellite trails, the intersecting portions of the trails are excluded from the analysis. The initial trail reconstruction process also excludes small bright spots, which may correspond to unflagged cosmic rays or other artifacts. Following this initial reconstruction and removal step, pixels with values exceeding 1.5 times the root mean square (RMS) of the background noise and the local trail level are identified and replaced with the corresponding values from the initial trail reconstruction. An additional iteration of trail reconstruction is subsequently performed to refine the trail profile and achieve a more accurate restoration. The source selection procedure follows the criteria previously established. A convolution kernel requiring a minimum of 13 connected pixels is applied, and the SNR is measured within 2.5 times the Kron radius. A lower size threshold is set to exclude small spurious sources that were not identified by the pipeline. Sources with an SNR greater than 5 are retained for further analysis. The numbers of sources detected per image (129.4 square arcminutes) are 2001, 2456, 3496, 3753, 4168, 3866, and 2080 in the NUV, $u$, $g$, $r$, $i$, $z$, and $y$ bands, respectively. To evaluate the effectiveness of trail correction, changes in forced photometric parameters are compared between uncontaminated images and those affected by satellite trails. Parameter measurements are obtained using forced photometry within 2.5 times the Kron radius. We employ the Python library SEP to detect contaminated sources and compute the error in the measured parameters \citep{Barbary2016}.

%the sources number in a image is each band is that 
% NUV,    u,       g,       r,     i,      z,       y
%2001.2  2456.2  3496.3  3753.4  4167.8 3865.5   2079.65

\section{Result and discussion}

\subsection{Photometric Error Analysis}

Forced photometry results are used to assess the impact of LEOsat trails under different orbital altitudes and brightness models, as well as to evaluate the performance of trail reconstruction and correction procedures. Photometric measurements obtained after trail removal are compared with those from uncontaminated images to quantify the residual effects of satellite-induced contamination. For the multi-band analysis, the corresponding error distributions under different brightness conditions are quantified and summarized in Table 2. Figure 2 visualizes these results, comparing the outcomes with and without trail correction. For instance, taking V$_{\rm M}$ = 7 at an altitude of 550 km, the reductions in the NUV, $u$, $g$, $r$, $i$, $z$, and $y$ bands are approximately 54\%, 44\%, 38\%, 41\%, 39\%, 43\%, and 38\%, respectively. Similarly, for V$_{\rm M}$ = 7 at an altitude of 1200 km, the corresponding reductions are approximately 55\%, 39\%, 18\%, 27\%, 31\%, 31\%, and 57\%.

\begin{table}[H]
\center
\caption{Proportion of sources contaminated (as a function of magnitude error threshold) by LEOsats at different altitudes and brightness models in each CSST band}
\begin{tabular}{lllllllll}
\hline
& Mag error (V$_{\rm M}$ \& Altitude)   &NUV &$u$   &$g$  &$r$  &$i$  &$z$  &$y$   \\
\hline
&$\ >$ 0.01 (V$_{\rm M}$ = 7 \& 550 km)    &0.056$\%$ &0.064$\%$ &0.086$\%$ &0.097$\%$ &0.089$\%$ &0.10$\%$ &0.088$\%$  \\
&$\ >$ 0.01  (V$_{\rm M}$ = 7 \& 1200 km)   &0.086$\%$ &0.092$\%$ &0.12$\%$ &0.15$\%$  &0.16$\%$  &0.16$\%$  &0.23$\%$   \\
&$\ >$ 0.01  (V$_{\rm M}$ = 5 \& 550 km)   &0.068$\%$ &0.083$\%$ &0.13$\%$ &0.15$\%$  &0.13$\%$   &0.16$\%$ &0.15$\%$   \\
&$\ >$ 0.01  (V$_{\rm M}$ = 5 \& 1200 km)   &0.090$\%$ &0.12$\%$ &0.22$\%$  &0.27$\%$  &0.28$\%$  &0.30$\%$  &0.24$\%$   \\
\hline
&$\ >$ 0.01 (V$_{\rm M}$ = 7 \& 550 km repair)    &0.026$\%$ &0.036$\%$ &0.053$\%$ &0.057$\%$ &0.054$\%$ &0.057$\%$ &0.055$\%$  \\
&$\ >$ 0.01  (V$_{\rm M}$ = 7 \& 1200 km repair)   &0.039$\%$ &0.056$\%$ &0.099$\%$ &0.11$\%$  &0.11$\%$  &0.11$\%$  &0.098$\%$   \\
&$\ >$ 0.01  (V$_{\rm M}$ = 5 \& 550 km repair)   &0.039$\%$ &0.048$\%$ &0.064$\%$ &0.072$\%$ &0.069$\%$ &0.075$\%$ &0.074$\%$   \\
&$\ >$ 0.01  (V$_{\rm M}$ = 5 \& 1200 km repair)   &0.061$\%$ &0.079$\%$ &0.13$\%$  &0.14$\%$  &0.14$\%$  &0.14$\%$  &0.14$\%$   \\
\hline
\end{tabular}
\caption*{\small{The upper rows correspond to results without trail removal, while the lower rows show results after LEOsat trail reconstruction and removal ('repair').}}
\end{table}

\begin{figure}
\centering
\includegraphics[width=0.48\linewidth]{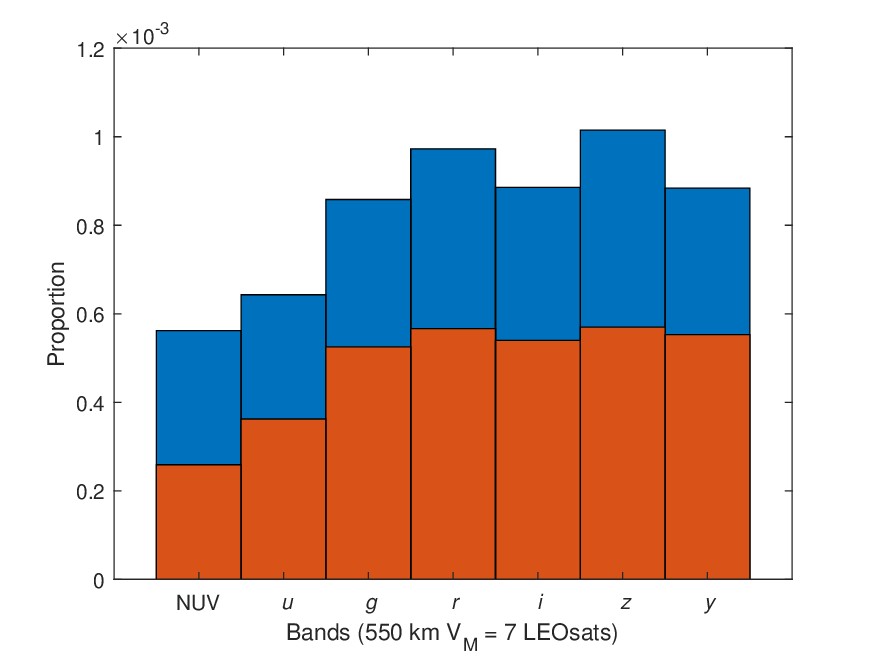}
\label{fig:side:a}
\hspace{0.01\linewidth}
\includegraphics[width=0.48\linewidth]{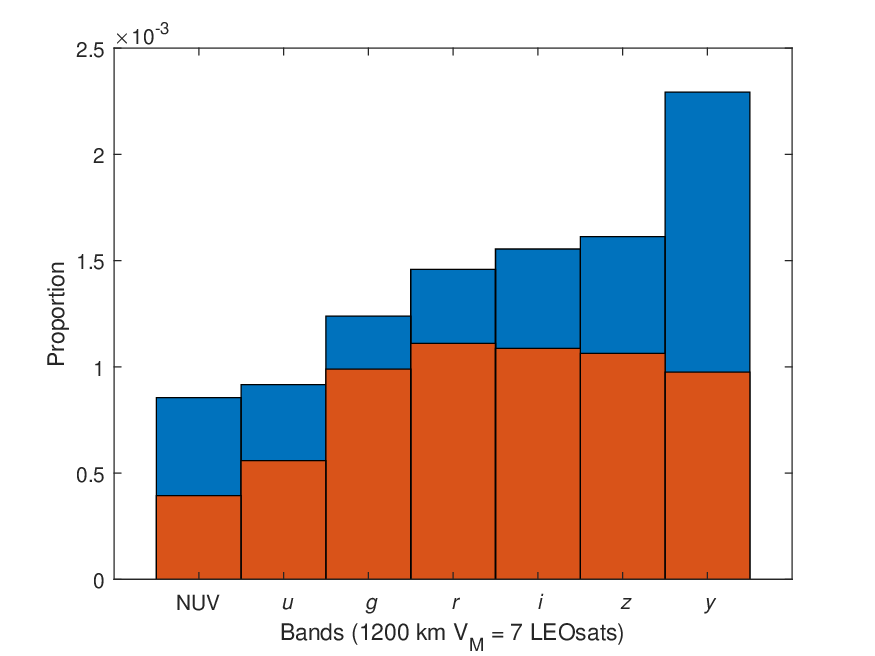}
\label{fig:side:b}
\includegraphics[width=0.48\linewidth]{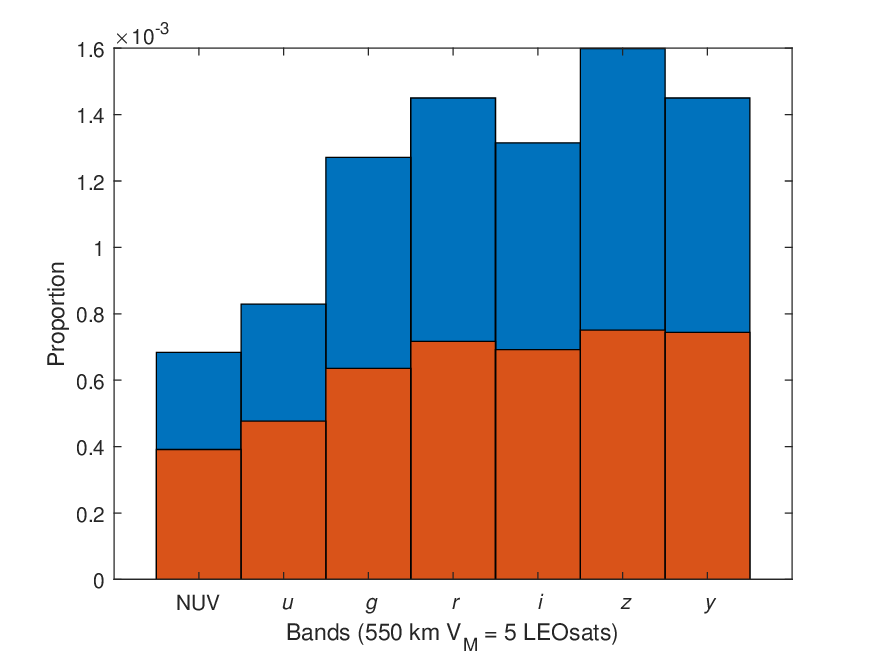}
\label{fig:side:c}
\hspace{0.01\linewidth}
\includegraphics[width=0.48\linewidth]{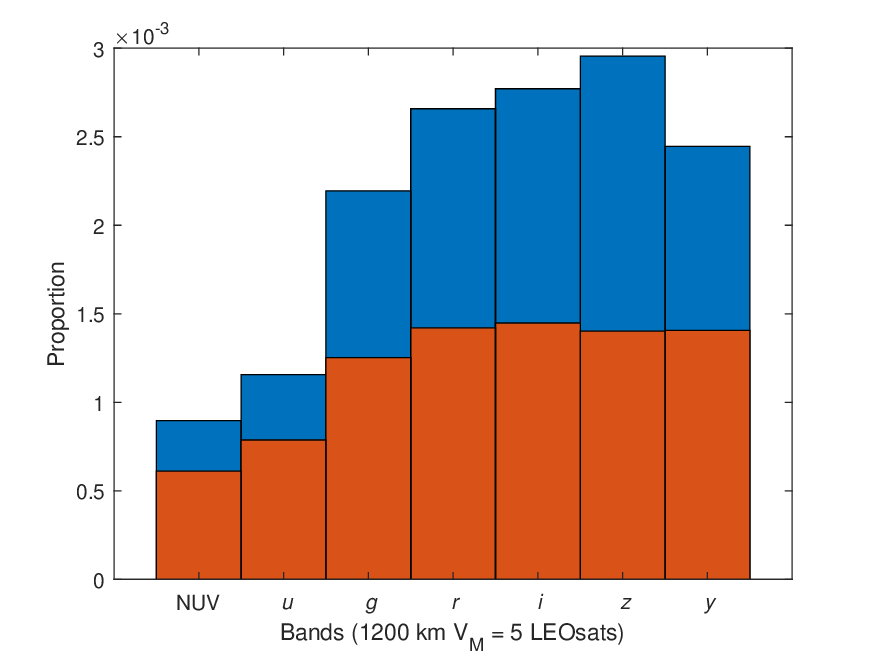}
\label{fig:side:d}
\caption{ 
The figure illustrates the results summarized in Table 2, comparing the outcomes with and without trail repair. The magnitude error threshold is 0.01; blue denotes the results without trail repair, and red represents the trail-repaired case. The left panels show the results for satellites at an altitude of 550 km, while the right panels correspond to an altitude of 1200 km. The upper panels present the results for V$_{\rm M} = 7$, and the lower panels for V$_{\rm M} = 5$.}
\end{figure}

A comparison is made between the brighter $i$-band and the relatively fainter NUV band. In Figure 3, we use the $i$-band relative photometric error induced by LEOsats on contaminated sources to represent the contamination effect. It is evident that the photometric error decreases as the distance from the satellite trail increases. Figure 4 presents the corresponding results for the NUV band, allowing direct comparison with the $i$-band case. Analyses are conducted for each observational band at orbital altitudes of 550 km and 1200 km, assuming minimum apparent V-band magnitudes (V$_{\rm M}$) of 5 and 7. The NUV band of the LEOsat trail is considerably fainter than that of the $i$-band. Consequently, in the central regions of satellite trails the contamination effect is further diminished, resulting in an even lower fraction of significantly affected sources.

\begin{figure}
\centering
\includegraphics[width=0.48\linewidth]{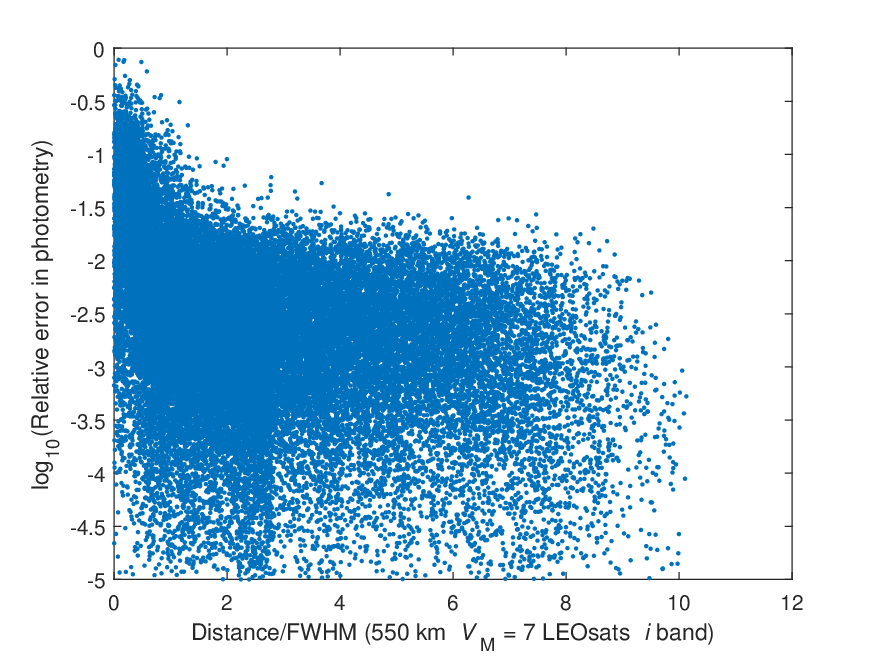}
\label{fig:side:a}
\hspace{0.01\linewidth}
\includegraphics[width=0.48\linewidth]{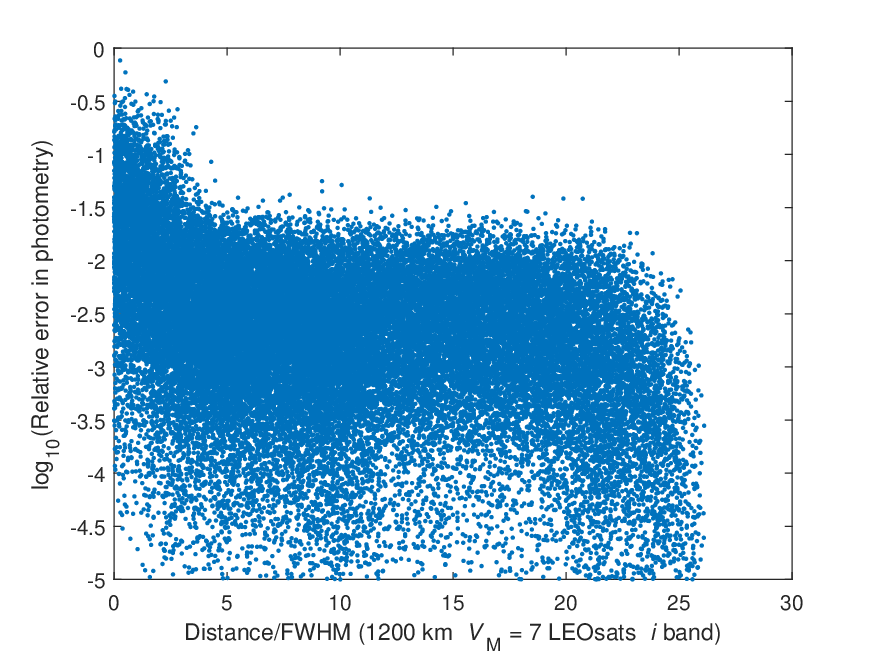}
\label{fig:side:b}
\includegraphics[width=0.48\linewidth]{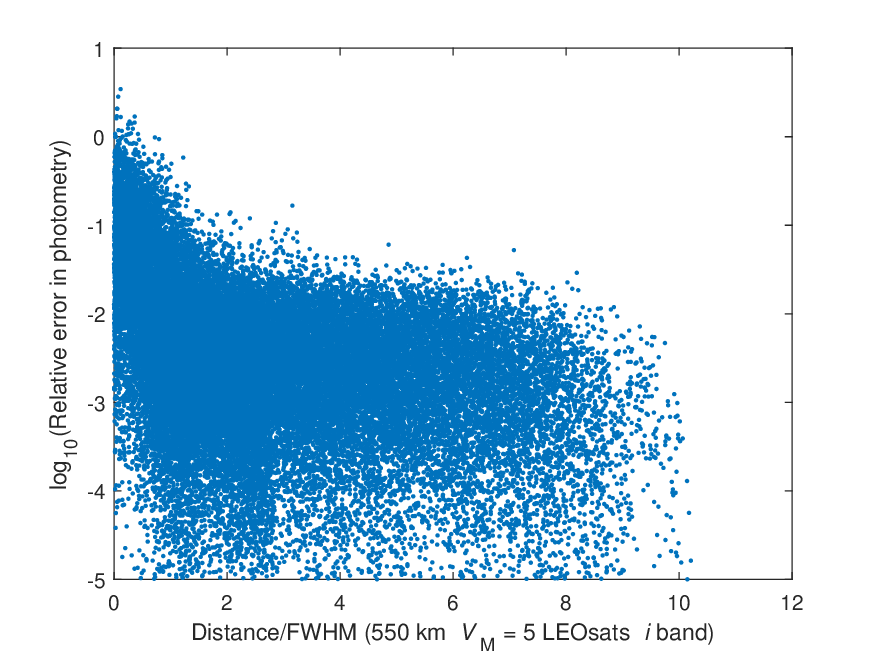}
\label{fig:side:c}
\hspace{0.01\linewidth}
\includegraphics[width=0.48\linewidth]{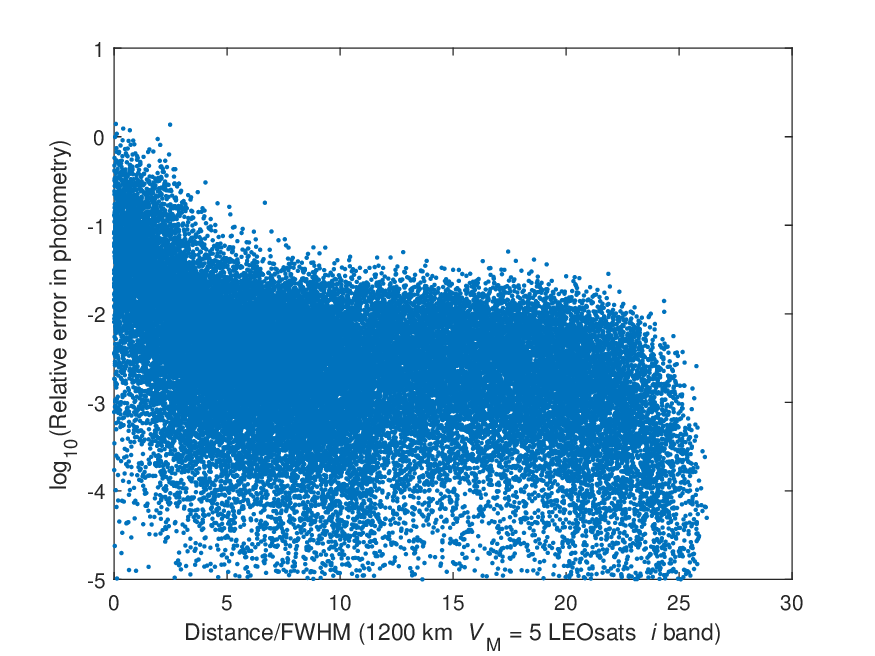}
\label{fig:side:d}
\caption{ 
The figure illustrates the relationship in the $i$-band between the distance of sources from LEOsat trails, normalized by the trail’s FWHM, and the decimal logarithm (log\textsubscript{10}) of the relative photometric error induced by the satellites. The left panels show results for satellites at an altitude of 550 km, while the right panels correspond to an altitude of 1200 km. The upper panels present results for V$_{\rm M} = 7$, and the lower panels for V$_{\rm M} = 5$.}
\end{figure}

\begin{figure}
\centering
\includegraphics[width=0.48\linewidth]{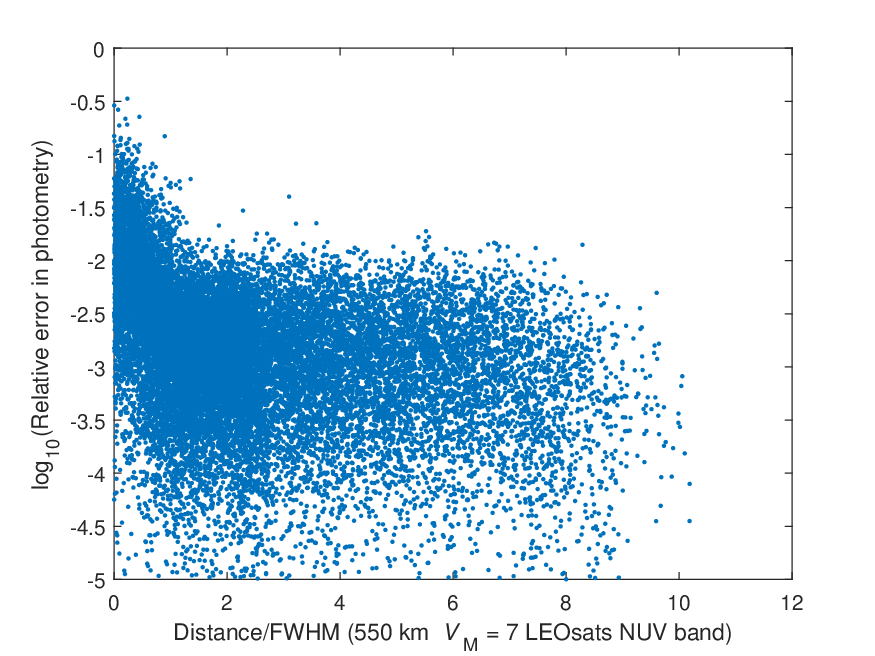}
\label{fig:side:a}
\hspace{0.01\linewidth}
\includegraphics[width=0.48\linewidth]{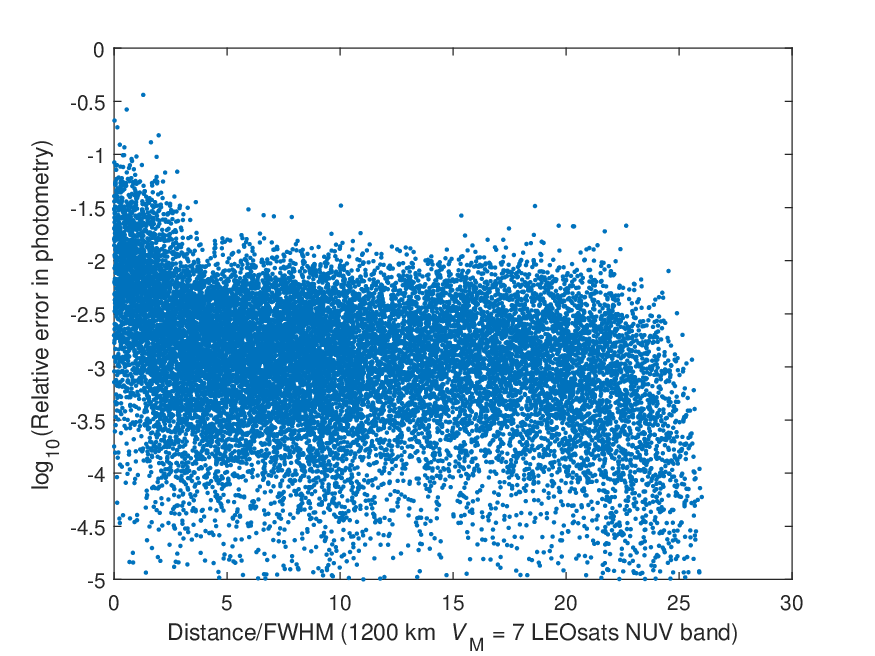}
\label{fig:side:b}
\includegraphics[width=0.48\linewidth]{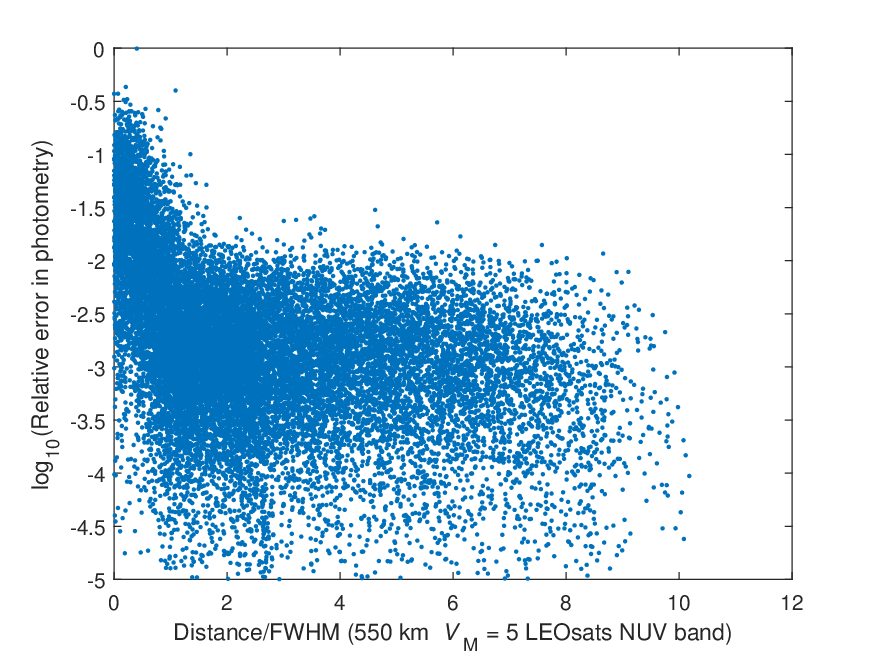}
\label{fig:side:c}
\hspace{0.01\linewidth}
\includegraphics[width=0.48\linewidth]{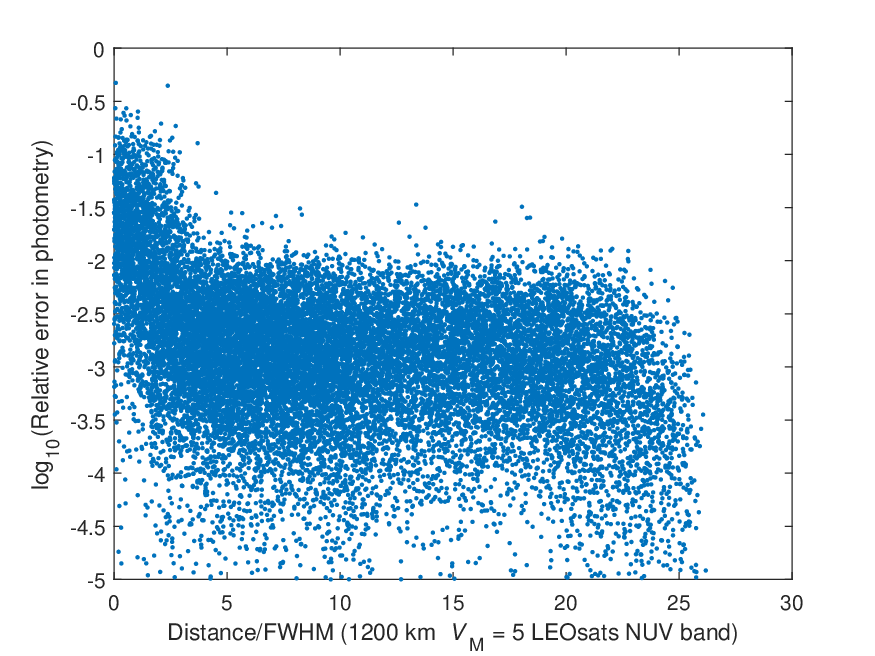}
\label{fig:side:d}
\caption{ 
The figure illustrates the relationship in the NUV-band between the distance of sources from LEOsat trails, normalized by the trail’s FWHM, and the decimal logarithm (log\textsubscript{10}) of the relative photometric error induced by the satellites. The panel layout is the same as in Figure 2. At the trail center, the distribution in the NUV band is narrower and lower than that in the $i$-band.}
\end{figure}

\subsection{Comparison with Other Noise Sources}

The SNR is closely related to the accuracy of source photometry, centroid determination, and shape measurements. The relative forced photometric error induced by LEOsat trails, multiplied by the source's SNR in the absence of the trail, defines an index for comparing photometric errors caused by satellite trail reconstruction residuals to those arising from other noise sources. This index, whose decimal logarithm is denoted as $y$-value, captures residuals comprising both photon noise from the trails and reconstruction errors. The contamination fraction is estimated by relating the expected trail lengths from a population of $10^5$ LEOsats at various altitudes to the total number of sources in the image. A value of $y$-value $ = -1$ indicates the trail-induced error is one-tenth that of other noise sources. Simulated images with and without reconstructed satellite trails are generated for each observational band to compare these effects. By analyzing relative forced photometric errors weighted by their SNRs, we quantify the fraction of sources for which the residual trail-induced error exceeds one-tenth of other noise contributions ($y$-value $ > -1$). These statistics, summarized in Table 3, provide a quantitative assessment of contamination across multiple bands. The distributions vary significantly, for example, the lower source density and fainter satellite trails in the NUV band lead to a smaller fraction of affected sources compared to the $i$-band. As shown in Figure 5, only a small proportion of sources remain significantly impacted after trail reconstruction, confirming the effectiveness of the correction process in reducing contamination. Consistent with the above results, trail repair leads to a notable reduction in the fraction of significantly affected sources (defined as those with $y$-values greater than –1). For example, with V$_{\rm M}$ = 7 at an altitude of 550 km, the reductions in the NUV, $u$, $g$, $r$, $i$, $z$, and $y$ bands are approximately 51\%, 42\%, 39\%, 44\%, 38\%, 44\%, and 44\%, respectively. At a higher altitude of 1200 km, the corresponding reductions are approximately 56\%, 40\%, 29\%, 35\%, 33\%, 39\%, and 73\%.

\begin{table}[H]
\center
\caption{Proportion of contaminated sources with LEOsat trail-induced errors after reconstruction exceeding one-tenth of other noise contributions}
\begin{tabular}{lllllllll}
\hline
& $y$-value (V$_{\rm M}$ \& Altitude)   &NUV &$u$   &$g$  &$r$  &$i$  &$z$  &$y$   \\
\hline
&$\ >$ -1 (V$_{\rm M}$ = 7 \& 550 km)  &0.059$\%$ &0.067$\%$ &0.093$\%$ &0.11$\%$ &0.098$\%$ &0.11$\%$ &0.091$\%$ \\
&$\ >$ -1 (V$_{\rm M}$ = 7 \& 1200 km) &0.089$\%$ &0.096$\%$ &0.14$\%$ &0.17$\%$  &0.18$\%$  &0.18$\%$  &0.27$\%$ \\
&$\ >$ -1 (V$_{\rm M}$ = 5 \& 550 km)  &0.070$\%$ &0.087$\%$ &0.15$\%$ &0.17$\%$ &0.15$\%$ &0.18$\%$ &0.15$\%$ \\
&$\ >$ -1 (V$_{\rm M}$ = 5 \& 1200 km) &0.091$\%$ &0.12$\%$  &0.26$\%$  &0.32$\%$ &0.33$\%$   &0.35$\%$  &0.26$\%$  \\
\hline
&$\ >$ -1 (V$_{\rm M}$ = 7 \& 550 km repair)  &0.029$\%$ &0.039$\%$ &0.057$\%$ &0.062$\%$ &0.061$\%$ &0.062$\%$ &0.051$\%$ \\
&$\ >$ -1 (V$_{\rm M}$ = 7 \& 1200 km repair) &0.039$\%$ &0.058$\%$ &0.099$\%$ &0.11$\%$  &0.12$\%$  &0.11$\%$  &0.083$\%$ \\
&$\ >$ -1 (V$_{\rm M}$ = 5 \& 550 km repair)  &0.042$\%$ &0.049$\%$ &0.073$\%$ &0.083$\%$ &0.081$\%$ &0.088$\%$ &0.073$\%$ \\
&$\ >$ -1 (V$_{\rm M}$ = 5 \& 1200 km repair) &0.064$\%$ &0.079$\%$ &0.14$\%$  &0.16$\%$ &0.17$\%$   &0.16$\%$  &0.13$\%$  \\

\hline
\end{tabular}
\caption*{\small{The upper rows correspond to results without trail removal, while the lower rows show results after LEOsat trail reconstruction and removal ('repair').}}
\end{table}

\begin{figure}[H]
\centering
\includegraphics[width=0.48\linewidth]{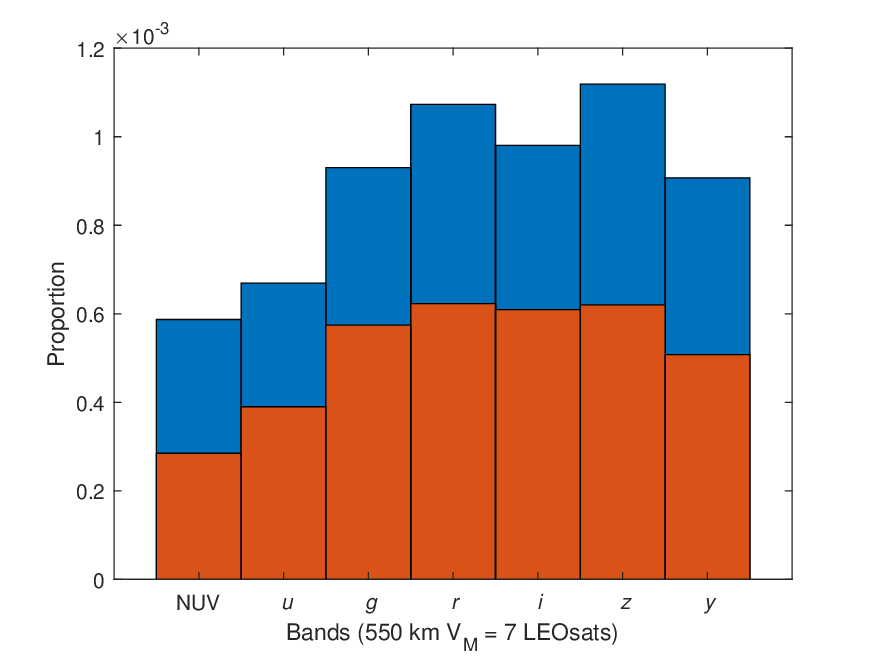}
\label{fig:side:a}
\hspace{0.01\linewidth}
\includegraphics[width=0.48\linewidth]{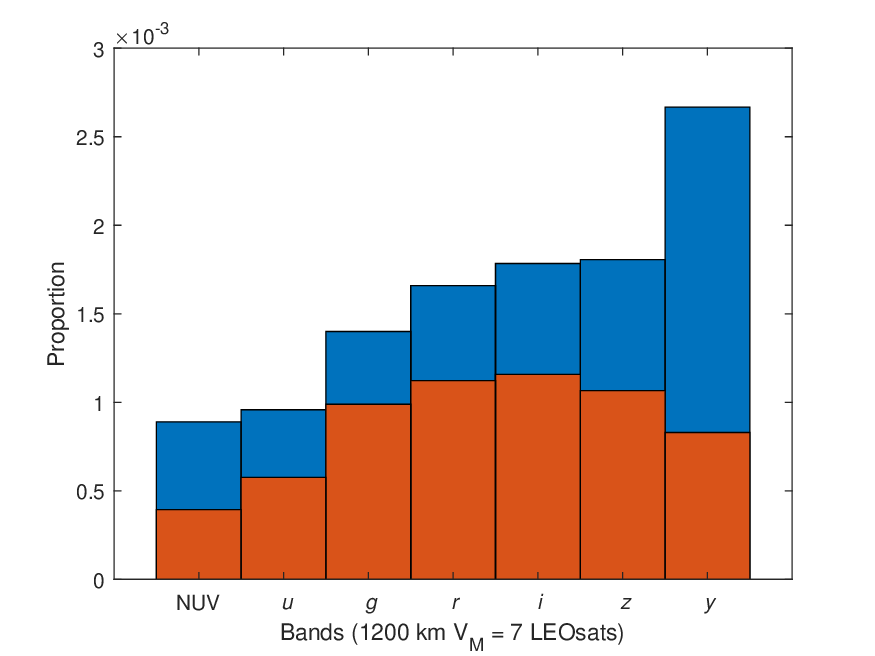}
\label{fig:side:b}
\includegraphics[width=0.48\linewidth]{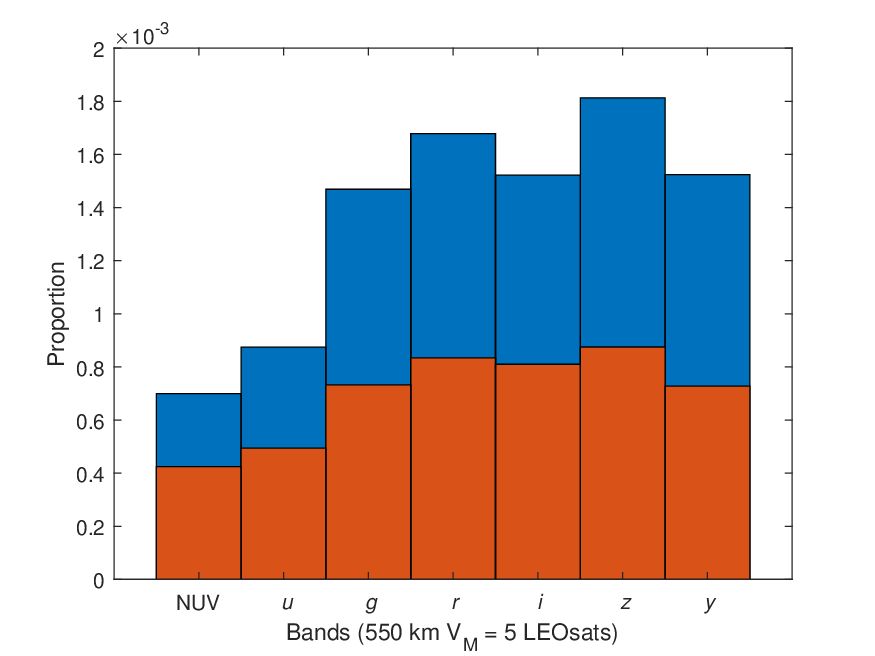}
\label{fig:side:c}
\hspace{0.01\linewidth}
\includegraphics[width=0.48\linewidth]{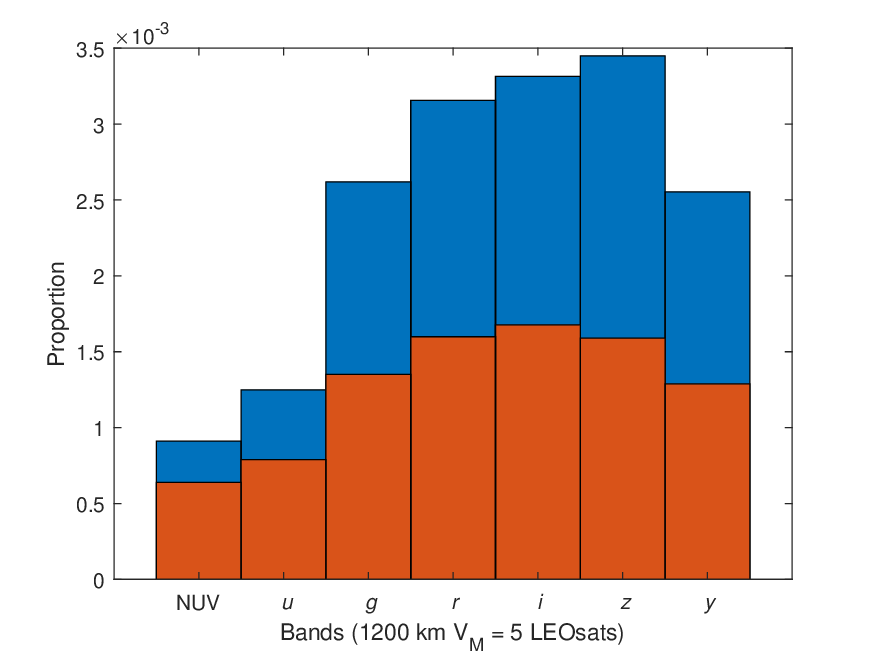}
\label{fig:side:d}
\caption{ 
The figure illustrates the results summarized in Table 3, comparing outcomes with and without trail repair, where blue denotes results without repair and red represents the trail-repaired case. It shows the proportion of contaminated sources whose LEOsat trail-induced errors after reconstruction exceed one-tenth of other noise contributions. The left panels correspond to satellites at 550 km altitude, while the right panels represent 1200 km altitude. The upper panels present results for V$_{\rm M} = 7$, and the lower panels for V$_{\rm M} = 5$.}
\end{figure}

In this study, simulations were conducted under the assumption that source positions are known, focusing primarily on photometric errors and noise variations. A full source extraction pipeline was not implemented, and as such, statistical metrics related to completeness and false detection rates were not evaluated. Moreover, considering that the CSST will perform multi-band and repeated exposures over the same sky region, the impact of source completeness and false positives in a single exposure on the final scientific measurements is expected to be limited. It is worth noting that the presented SNR comparisons are strongly correlated with other key metrics—such as centroid shifts and source recovery rates—and thus provide a meaningful indication of the overall impact of satellite trail contamination on source measurements.

\section{Conclusion}

For the evaluation of each photometric band in CSST, the overall fraction of affected sources remains low following satellite trail correction. Compared to our previous work, the simulated images used in this study more closely reflect a single real CSST observation and are no longer limited to a single-band analysis. This study employs CSST simulated multi-band images containing various realistic effects to assess the effectiveness of the proposed method under conditions comparable to real observations. After the reconstruction and removal of satellite trails, in bands corresponding to relatively bright trails and higher orbital altitudes, the proportion of sources with trail-induced errors exceeding one-tenth of other noise contributions is generally maintained at a very low level. In bands where satellite trails appear fainter, this proportion is even lower.

The results indicate a significant reduction in the fraction of sources falling into two categories: (1) those with magnitude errors greater than 0.01 mag attributable to LEOsats, and (2) those for which LEOsat-induced noise exceeds 10\% of the total noise contribution from other sources. Following trail repair, the analysis reveals a reduction of over 50\% in the fraction of affected sources in the NUV band at both 550 km and 1200 km altitudes, assuming a maximum brightness of 7 in the V band. In the $i$ band, the reduction exceeds 30\%. The reconstruction and correction of LEOsat trails across multiple bands provide valuable guidance for mitigating directional contamination from LEOsats in future CSST observations. The multi-band simulated images incorporate numerous realistic instrumental and observational effects, and as such, some residual contamination may still persist after correction. Notably, bright astronomical sources and unflagged cosmic rays intersecting with satellite trails can compromise the accuracy of reconstruction. During the correction process, these bright features must be identified and either replaced or mitigated to reduce their impact on the satellite trail reconstruction.

Due to brightness differences between photometric bands, the level of contamination varies, with fainter bands generally being less affected. Moving beyond conventional masking strategies that discard contaminated regions, this work explores an alternative approach by reconstructing and subtracting satellite trail profiles, aiming to preserve as much valid observational data as possible. Improving the quality of trail reconstruction is therefore essential for minimizing contamination and ensuring data integrity in large-scale space-based survey programs.

\begin{acknowledgements}
This work was supported by National Key R\&D Program of China No.2022YFF0503400. This work is supported by the China Manned Space Program with grant no.CMS-CSST-2025-A20. This work is based in part on the mock data created by the CSST Simulation Team, which is supported by the CSST scientific data processing and analysis system of the China Manned Space Project.
\end{acknowledgements}
%This work is based [in part] on the mock data [or the softwares] created by the CSST Simulation Team, which is supported by the CSST scientific data processing and analysis system of the China Manned Space Project. <someone> acknowledges the support by <Program Name> (No. <Program Number>).

%\appendix                  %%appendicial material is supported

%\section{This shows the use of appendix}
%A postscript file is actually an ASCII text file (you may even edit it).
%However, you need to transfer a PDF file or any compressed or packaged
%file in binary mode when using FTP.

%\section{What is SCI?}
%SCI is the abbreviation of Science Citation Index system powered by
%the Institute for Scientific Information (ISI). For details please
%visit {\it http://apps.isiknowledge.com}.

\label{lastpage}
\end{document}